\documentstyle[preprint,eqsecnum,aps]{revtex}

\tightenlines
\begin{document}
\draft
\preprint{HEP/123-qed}
\title{An importance sampling algorithm
for generating exact eigenstates of the nuclear Hamiltonian}
\author{F. Andreozzi, N. Lo Iudice\cite{ema}, and A. Porrino}
\address{Dipartimento di Scienze Fisiche, Universit\`{a} di
Napoli Federico
II,\\
and Istituto Nazionale di Fisica Nucleare. \\
Complesso Universitario di Monte S. Angelo, Via Cinthia, 80126
Napoli, Italy
}%
\date{\today}
\maketitle
\begin{abstract}
We endow a recently devised algorithm for generating exact
eigensolutions of large matrices with an importance sampling,
which is in control of the extent and accuracy of the truncation
of their dimensions. We made several tests on typical nuclei using
a correlated basis obtained from partitioning the shell model
space. The sampling so implemented allows not only for a
substantial reduction of the shell model space but also for an
extrapolation to exact eigenvalues and E2 strengths.
\end{abstract}
\narrowtext

\section{Introduction}

New effective methods for solving exactly the shell model (SM)
eigenvalue problem in complex nuclei have been developed in recent
years. A notable one is the Monte Carlo technique \cite{MC} for
generating exact SM ground states \cite{Jon92,Orm94,White00}. Its
application, however, has been rather restricted because of the
well known {\it minus-sign} problem.

Alternative methods face directly the diagonalization of the SM
Hamiltonian by resorting to  algorithms like Lanczos \cite{Lanc}
and Davidson \cite{Dav}. The critical points of direct
diagonalization methods are the amount of memory needed and the
time spent in the diagonalization process.

In order to overcome these limitations there have been attempts to
combine the stochastic methodology  with the standard
diagonalization approach
\cite{Var94,Horo94,Horo99,Horo03,MC1,MC2}. In \cite{Var94} the use
of Gaussian-like single-particle basis states with random
oscillator frequencies was suggested. The authors of
\cite{Horo94,Horo99,Horo03} proposed a stochastic truncation of
the SM matrices and suggested for the energies so obtained an
exponential extrapolation law to the exact eigenvalues. In the so
call quantum Monte Carlo diagonalization method
(QMCD)\cite{MC1,MC2} the reduction of the dimension of the SM
space is achieved by using the auxiliary field Monte Carlo
technique to select the {\it relevant} basis states. Though quite
successful, the latter method  is not free from problems. Indeed,
one has to deal with the redundancy of the basis states, which may
slow considerably the convergence of the procedure,  as well as
with the problem of restoring the symmetries generally broken in
stochastic approaches.

In a recent paper \cite{Andre02}, we have developed an iterative
algorithm for determining a selected set of eigenvectors of a
large matrix which is faster than the other, currently adopted,
algorithms, including Lanczos, and extremely simple to be
implemented. It is, moreover, {\it robust}, yielding always {\it
ghost}-free stable solutions.

Like the other methods, however, it requires the storage of at
least one eigenvector, which exceeds the limits of modern
computers in many complex systems. The present paper deals with
the problem of overcoming this limitation. To this purpose, we
endow the algorithm with an importance sampling for reducing the
sizes of the matrix. The sampling is closely linked to the
iterative algorithm and is in full control of the accuracy of the
eigensolutions. This will emerge clearly from the results of the
exhaustive tests presented and discussed in this paper. In fact,
we adopted the importance sampling to solve the shell model
problem for three typical nuclei, the semi-magic $^{108}$Sn, the
$N=Z$ doubly even $^{48}$Cr, and the $N \neq Z$ odd $^{133}$Xe. In
order to enhance its efficiency  we used a correlated basis
obtained from partitioning the shell model space according to a
method developed in Ref. \cite{Andre01}. We will show that the
method not only allows for a drastic truncation of the SM space
but yields naturally extrapolation laws to exact eigenvalues and
eigenvectors  as well as to the occupation numbers and the $E2$
transition probabilities.

\section{The algorithm}
Let us first give a brief outline of the algorithm \cite{Andre02}.
For the sake of simplicity, we consider here a symmetric matrix
\begin{equation}
A=\{a_{ij}\} = \{\langle i \mid \hat{A} \mid j \rangle\}
\end{equation}
representing a self-adjoint operator $\hat A$ in an orthonormal
basis $\{\mid 1 \rangle,\mid 2 \rangle,\dots,\mid N \rangle \}$.
The algorithm goes through  several iteration  loops. The first
loop  consists of the following steps:

\noindent 1a) Diagonalize the two-dimensional matrix $(a_{ij})$
(i,j=1,2),

\noindent 1b) select the lowest eigenvalue $\lambda_{2}$  and the
corresponding eigenvector
\begin{equation}
\mid \phi_{2} \rangle =
 c^{(2)}_{1} \mid 1 \rangle +  c^{(2)}_{2 }\mid 2
\rangle,
\end{equation}
1c) for  $j = 3, \dots, N$, diagonalize the matrix
\begin{eqnarray}
\left( \begin{array}{cc} \lambda_{j-1} & b_j \\ b_{j}
& a_{jj}
\end{array} \right)
\label{matrj}
\end{eqnarray}
where $b_j =
\langle \phi_{j-1} \mid \hat A \mid j \rangle$
and select the lowest eigenvalue $\lambda_j$ together with the
corresponding eigenvector $\mid \phi_j \rangle$.
This zero approximation loop yields the approximate eigenvalue
and eigenvector
\begin{equation}
E^{(1)}  \equiv \lambda_N ,\,\,\,\,\,\,\,\,\mid \psi^{(1)} \rangle \equiv
\mid \phi_N \rangle  =
\sum_{i=1}^{N} c^{(N)}_{i} \mid i \rangle \label{psi}.
\end{equation}
With these new entries
we start an iterative procedure which goes through
$n = 2,3, \dots \;\;\;$ refinement loops, consisting
of the same steps with  the following modification.
At each step  $j = 1,2, \dots, N$ of the $n$-th loop ($n>1$)
we have to solve  an eigenvalue problem  of
general form, since the states $\mid
\phi_{j-1}  \rangle$ and $\mid j \rangle$ are no longer orthogonal.
The  eigenvalue
$E^{(n)} \equiv \lambda_N $ and
eigenvector $\mid \psi^{(n)} \rangle  \equiv \mid \phi_{N}
\rangle$ obtained after the $n$-th loop are  proven to converge
to the exact eigenvalue $E$ and eigenvector
$\mid \psi \rangle$ respectively \cite{Andre02}.

The algorithm has been shown to be completely equivalent to the
method of optimal relaxation \cite{Shavitt} and has therefore a
variational foundation.

Because of its matrix formulation, however, it can be generalized
with minimal changes so as to generate at once an arbitrary number
$v$ of eigensolutions. Indeed, the first loop goes  through the
following steps:

\noindent 1a) Start with  $m(\geq v)$ basis vectors
 and diagonalize the $m$-dimensional principal submatrix $\{a_{ij}\}\,
 (i,j=1,m)$,

\noindent 1b) select the $v$ lowest eigenvalues $ \lambda_1,
\lambda_2, \dots ,\lambda_{v}$ and the corresponding eigenvectors
$\mid \phi_1 \rangle, \mid \phi_2 \rangle, \dots , \mid \phi_{v}
\rangle.$

\noindent For $k = 2,3, \dots, k_{N}$, where $k_{N}$ are the steps
necessary to exhaust the whole $N$-dimensional matrix,

\noindent 1c) diagonalize the matrix
\begin{equation}
\left( \begin{array}{cc} \Lambda _{k} & B_{k}\\
 B_{k}^{T} & A_{k} \end{array} \right), \label{Amult}
\end{equation}
where $\Lambda _{k}$ is a $v$-dimensional diagonal matrix whose
non-zero entries are the eigenvalues $ \lambda_1^{(k-1)},
\lambda_2^{(k-1)}, \dots ,\lambda_{v}^{(k-1)}$,
$A_{k}=\{a_{ij}\}\,(i,j = (k-1)p +1, \dots,kp)$ is a
$p$-dimensional submatrix, $B_{k}$ and its transpose are matrices
composed of the matrix elements $b_{ij}^{(k)} = \langle
\phi_{i}^{(k-1)} \mid \hat A \mid j \rangle$ (i = 1, $\dots   ,
v;\, j= (k-1)p +1,\dots, kp$).

\noindent 1d) Select the $v$ lowest eigenvalues $
\lambda_1^{(k)},\lambda_2^{(k)}, \dots ,\lambda_{v}^{(k)}$ and the
corresponding eigenvectors $\mid \phi_1^{(k)} \rangle, \mid
\phi_2^{(k)} \rangle, \dots , \mid \phi_{v}^{(k)} \rangle.$

Once the basis is exhausted, the process  yields $v$ approximate
eigenvalues and eigenvectors, which are the new entries for a new
iteration. This goes through the same steps with one essential
modification. Each loop, in fact, can be viewed as the solution of
the eigenproblem for the restriction ${\hat A}|_S$ of the operator
$\hat A$ to a subspace defined by the span of the set  of vectors
$\mid i_{p}\rangle \equiv\{ \phi_{1}^{(k)}, \dots ,
\phi_{v}^{(k)}, \mid(k-1)p+1 \rangle, \dots ,\mid kp \rangle \}$.
Since the vectors $\phi_{i}^{(k)}$ are linear superpositions of
all the basis vectors $\mid j \rangle$ we started with, the basis
obtained after the first approximation loop is no longer
orthonormal, just as in the one-dimensional eigenspace, and may be
even redundant. We have therefore to solve an eigenvalue problem
of general form. This is done most effectively through a pivotal
Choleski decomposition of the overlap matrix $\langle i_{p} \mid
 i'_{p}\rangle $ \cite{Cov}. As outlined in a forthcoming
paper, this procedure is very easy to be implemented and extremely
fast, specially in view of the very limited dimensions of the
matrices coming into play. With this modifications, the subsequent
iteration loops proceed as the first one. In this way we generate
a sequence of $v$  vectors $\psi^{(k)}_1, \dots, \psi^{(k)}_{v}$.
The restriction of the operator $\hat A$ to these sets defines a
sequence of diagonal matrices, whose non zero elements are the
current eigenvalues $\lambda^{(k)}_1, \dots , \lambda^{(k)}_{v}$,
with decreasing norms. This monotonic sequence is certainly
bounded from below and therefore convergent.

\section{Importance sampling}
The just outlined algorithm, though of simple implementation,
requires
the storage of at least one eigenvector. Since for many complex systems
the dimensions of the Hamiltonian matrix become prohibitively large,
one must rely on some importance sampling which allows for
a truncation of the space by
selecting only the basis states  relevant to
the exact eigensolutions. A notable example
is the stochastic diagonalization method \cite{Dav}, which
samples the basis states relevant to
the ground state through a combination of plane (Jacobi) rotations and
matrix inflation.

A similar, but more efficient, strategy can be implemented in the
framework of our diagonalization process. Exploiting the fact that
the algorithm  yields quite accurate solutions
already in the first approximation loop,
we can devise a sampling which makes use of the first
loop only.
This is to be accordingly modified and goes through the following steps:

1a) Turn the $v$-dimensional principal submatrix $\{a_{ij}\}\,
 (i,j=1,v)$ into the diagonal form $\Lambda_v$with eigenvalues
 $\lambda_1, \lambda_2, \dots, \lambda_v$.

 1b) For $j = v+1,\dots, N$,  diagonalize the
 $v+1$-dimensional matrix

\begin{equation}
A= \left( \begin{array}{cc} \Lambda_{v} & \vec{b}_{j}\\
 \vec{b}_{j}^{T} & a_{jj} \end{array} \right), \label{Amult1}
\end{equation}
where $\vec{b}_{j}= \{b_{1j}, b_{2j}, \cdot, b_{vj}\}$.

1c) Select the lowest $v$ eigenvalues $\lambda^{\prime}_{i},
(i=1,v)$ and accept the new  state only if
\begin{equation}
\sum_{i=1,v} \mid \lambda^{\prime}_{i} - \lambda_{i}\mid >
\epsilon \label{sampeps}
    \end{equation}
Otherwise restart from point 1b) with a new $j$.

We can avoid such a time consuming sampling procedure by resorting
to an alternative, though completely equivalent, route based on
the method developed for deriving an exact non perturbative shell
model Hamiltonian \cite{Andre96}. We carry the similarity
transformation
\begin{equation}
    A^{\prime} = \Omega^{-1} A \Omega
     \end{equation}
where
\begin{equation}
\Omega = \left( \begin{array}{cc}  I_{v} &  0\\
 \vec{\omega} & I_{Q} \end{array} \right). \label{Amult2}
\end{equation}
Here, $I_{v}$ is   the $v$-dimensional unit matrix and $\omega$ a
$v$-dimensional vector.

The transformed matrix has the following structure
\begin{equation}
A^{\prime}= \left( \begin{array}{cc} \Lambda_{v} +
\vec{b}_{j}\otimes \vec{\omega}& \vec{b}_{j}\\
 \vec{b}_{j}^{\prime} & a_{jj} - \vec{\omega} \cdot \vec{b}_{j}
\end{array} \right) \label{Amult3}
\end{equation}
where
\begin{equation}
\vec{b}_{j}^{\prime}= - (\vec{\omega} \cdot \vec{b}_{j})
 \vec{\omega} - \vec{\omega} \Lambda_{v} + a_{jj} \vec{\omega} +
\vec{b}_{j}^{T}. \label{bp0}
\end{equation}
We now  impose the decoupling condition
\begin{equation}
\vec{b}_{j}^{\prime}= - (\vec{\omega} \cdot \vec{b}_{j})
 \vec{\omega} - \vec{\omega} \Lambda_{v} + a_{jj} \vec{\omega} +
\vec{b}_{j}^{T} = 0. \label{bp}
\end{equation}
Once we find the solution $\omega$, the matrix element $a_{jj} -
\vec{\omega} \cdot \vec{b}_{j}$ becomes an eigenvalue of the
matrix $A^{\prime}$ and, therefore, of matrix $A$. Right
multiplying Eq. (\ref{bp}) by $\vec{b}_{j}$, we obtain the
dispersion relation
\begin{equation}
\vec{\omega} \cdot \vec{b}_{j}= - \sum_{i=1,v}
\frac{b^{2}_{ij}}{a_{jj} - \lambda_{i} - \vec{\omega} \cdot
\vec{b}_{j}}. \label{disprel}
\end{equation}
This admits $v+1$ solutions, corresponding to the $v+1$ eigenvalues
of A. In correspondence of the lowest solution
$(\vec{\omega} \cdot \vec{b}_{j})_{min}$, we get the maximum eigenvalue
\begin{equation}
 \lambda^\prime_{max} = a_{jj} - (\vec{\omega} \cdot
 \vec{b}_{j})_{min}.
\end{equation}
The eigenvalues $\lambda_{i}, i=1,\ldots,v$ separate at least in
weak sense the new eigenvalues $\lambda_{i}^\prime,
i=1,\ldots,v+1$ \cite{Wilk65}, namely
\begin{equation}
\lambda_{1}^{\prime} \leq \lambda_{1} \leq \lambda_{2}^{\prime}
\leq \lambda_{2}\leq \ldots \leq \lambda_{v}^{\prime} \leq
\lambda_{v} \leq \lambda_{max}^\prime.
\end{equation}
Since
\begin{equation}
\sum_{i=1,v+1} \lambda^{\prime}_{i} = \sum_{i=1,v}
\lambda^{\prime}_{i} + a_{jj} - (\vec{\omega} \cdot
\vec{b}_{j})_{min} = a_{jj} + \sum_{i=1,v}\lambda_{i}
\end{equation}
 we have
\begin{equation}
\sum_{i=1,v} (\lambda^{\prime}_{i} - \lambda_{i}) = (\vec{\omega}
\cdot \vec{b}_{j})_{min}.\label{samp1}
\end{equation}
Condition (\ref{sampeps}) is therefore equivalent to
\begin{equation}
 (\vec{\omega} \cdot \vec{b}_{j})_{min} > \epsilon. \label{samp2}
\end{equation}
The just outlined sampling procedure requires only the solution of
the dispersion equation (\ref{disprel}), which is of the type
\begin{equation}
  f(z) = z
\end{equation}
and fulfils the condition
\begin{equation}
1 - df(z)/dz >0.
\end{equation}
We can therefore easily solve it by using the Newton method of
derivative. This alternative sampling procedure is not only
rigorous but also more economical. It avoids, in fact, the
$(N-v)-$fold iterated diagonalization of $v+1$ dimensional
matrices.

\section{Multipartitioning method}
The  extent of truncation induced by the sampling is maximal when
the eigenvectors are highly localized. This condition is fulfilled
in most physical problems. Even when this is not the case, we can
approach the above condition by using a correlated basis obtained
by a multipartitioning method \cite{Andre01}. This goes through
the following prescriptions:

\noindent i) Partition the shell model space for N valence
nucleons into orthogonal subspaces, $P_{1}$ and $P_{2}$
according to
\begin{equation}
    P= P_{1} + P_{2},
\end{equation}
ii) distribute
$N_{1}$ and $N_{2}$ nucleons ($N_{1} + N_{2} = N$) among these subspaces
in all possible ways,

\noindent iii) decompose the Hamiltonian $H$ into
\begin{equation}
 H= H_{1} + H_{2} + H_{12},
 \end{equation}
 iv) solve the eigenvalue equations
\begin{equation}
H_{i}\mid\alpha_{i}N_{i}\rangle = E_{\alpha_i} \mid\alpha_{i}
N_{i}\rangle
\end{equation}
obtaining  the eigenstates $\mid \alpha_{1}N_{1}\rangle$ and $\mid
\alpha_{2}N_{2}\rangle$ of $H_{1}$ and $H_{2}$ respectively in
$P_{1}$ and $P_{2}$. Once this is done, it is possible to replace
the standard shell model basis with  one composed of the states
\begin{equation}
 \mid \alpha N \rangle = \mid \alpha_{1}N_{1} \alpha_{2} N_{2}\rangle.  \label{part1}
\end{equation}
We use the above basis to diagonalize the residual term $H_{12}$
of the SM Hamiltonian. The new basis is in general highly
correlated and, therefore, highly localized in the Fock space, a
feature which enhances considerably the efficiency of the method.

\section{Selective numerical tests}
We applied the sampling algorithm to the semi-magic $^{108}$Sn,
the N=Z even-even $^{48}$Cr and the $N>Z$ odd-even $^{133}$Xe. The
model spaces are:

\noindent 1) $P \equiv \{2d5/2,1g7/2,2d3/2,3s1/2,1h11/2\}$ for the
8 valence neutrons of $^{108}$Sn and for the 4 valence protons and
3 valence neutron holes of $^{133}$Xe,

\noindent 2)$P \equiv \{1f7/2,1f5/2,2p3/2,2p1/2\}$ for the 4
valence protons and neutrons of $^{48}$Cr.

We adopted a realistic effective interaction deduced from the
Bonn-A potential \cite{Mcl} for $^{108}$Sn and $^{133}$Xe, and
used the KB3 interaction \cite{KB3} for $^{48}$Cr. For $^{108}$Sn
we used the single particle (s.p.) energies (in MeV)
$\epsilon_{2d5/2}=0,\,\epsilon_{1g7/2}= 0.2,\,
\epsilon_{3s1/2}=2.2,\,\epsilon_{2d3/2}=2.3,\,\epsilon_{1h11/2}=2.9$.
Apart from the first two currently in use, the other energies were
deduced from a fit on $^{111}$Sn for $\{2d3/2,3s1/2\}$ and from
the level $9^{-}$ in $^{106}$Sn for $1h11/2$. We used the same set
of energies for protons in $^{133}$Xe, while for neutron-holes we
used the slightly different values
$\epsilon_{2d5/2}=0,\,\epsilon_{1g7/2}= 0.2,\,
\epsilon_{3s1/2}=1.72,\,\epsilon_{2d3/2}=1.88,\,\epsilon_{1h11/2}=2.7$.
The different set accounts effectively, at least in part, for the
significant asymmetry between proton and neutron numbers which
yield different contributions from the three- and more-body
forces, not included in the effective interaction. As for
$^{48}$Cr, we used the same energies adopted in Ref.\cite{KB3},
namely $\epsilon_{1f7/2}=0,\, \epsilon_{2p3/2}=
2.0,\,\epsilon_{2p1/2}= 4.0,\, \epsilon_{1f5/2}= 6.5$.

We partitioned the shell model space for $^{108}$Sn according the
following prescriptions:
\begin{eqnarray}
P \equiv \{2d5/2,1g7/2,2d3/2,3s1/2,1h11/2\}
 \begin{array}{ccc}  \nearrow P_{1}& \equiv& \{2d5/2,1g7/2\} \\
& &+\\\searrow P_{2}&\equiv &\{2d3/2,3s1/2,1h11/2\}.
\end{array}
\label{Part}
\end{eqnarray}
This partition is dictated by the large
energy gap ($\sim 2$ MeV) between the two corresponding sets of
single particle energies.

For  $^{48}$Cr and $^{133}$Xe, we simply decompose the space into
a proton and neutron subspace
\begin{equation}
 P = P_{p} + P_{n}.
\end{equation}
We adopted then the multipartitioning method \cite{Andre01} to
generate a new correlated basis
\begin{equation}
\mid j \rangle = \mid \alpha N \rangle
\end{equation}
and used this new basis to implement the importance sampling.
 
\subsection{Eigenvalues}
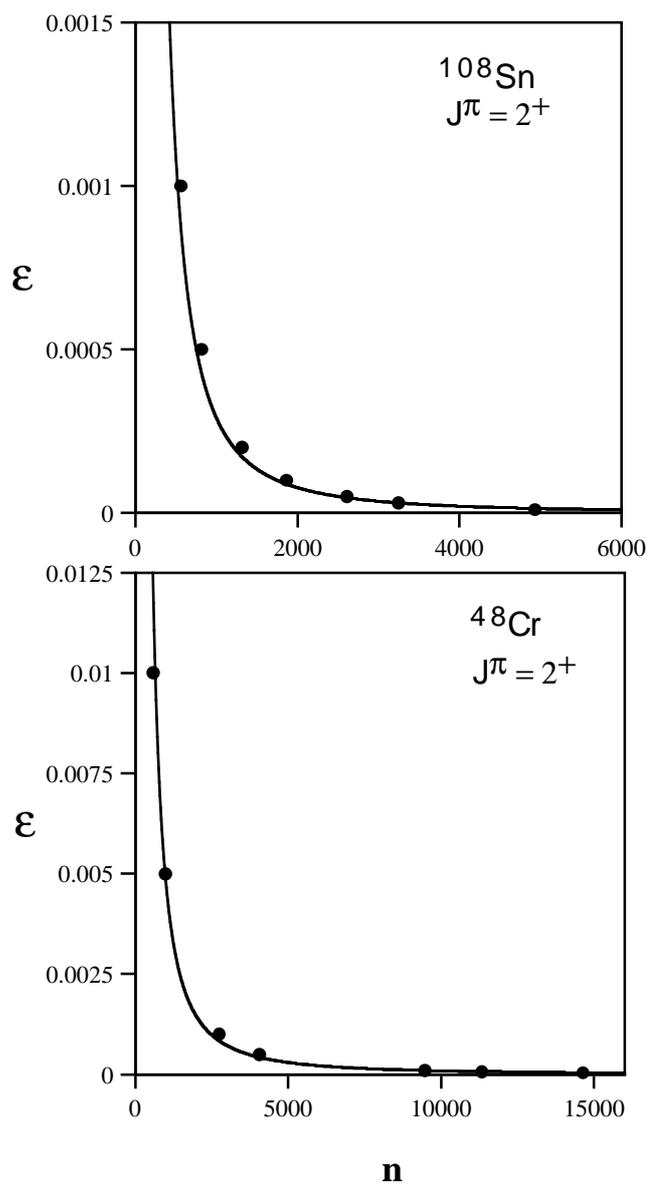
\begin{figure}
\caption{\label{fig1} Importance sampling parameters versus the
dimensions $n$ of the truncated matrices resulting from the
sampling.}
\end{figure}

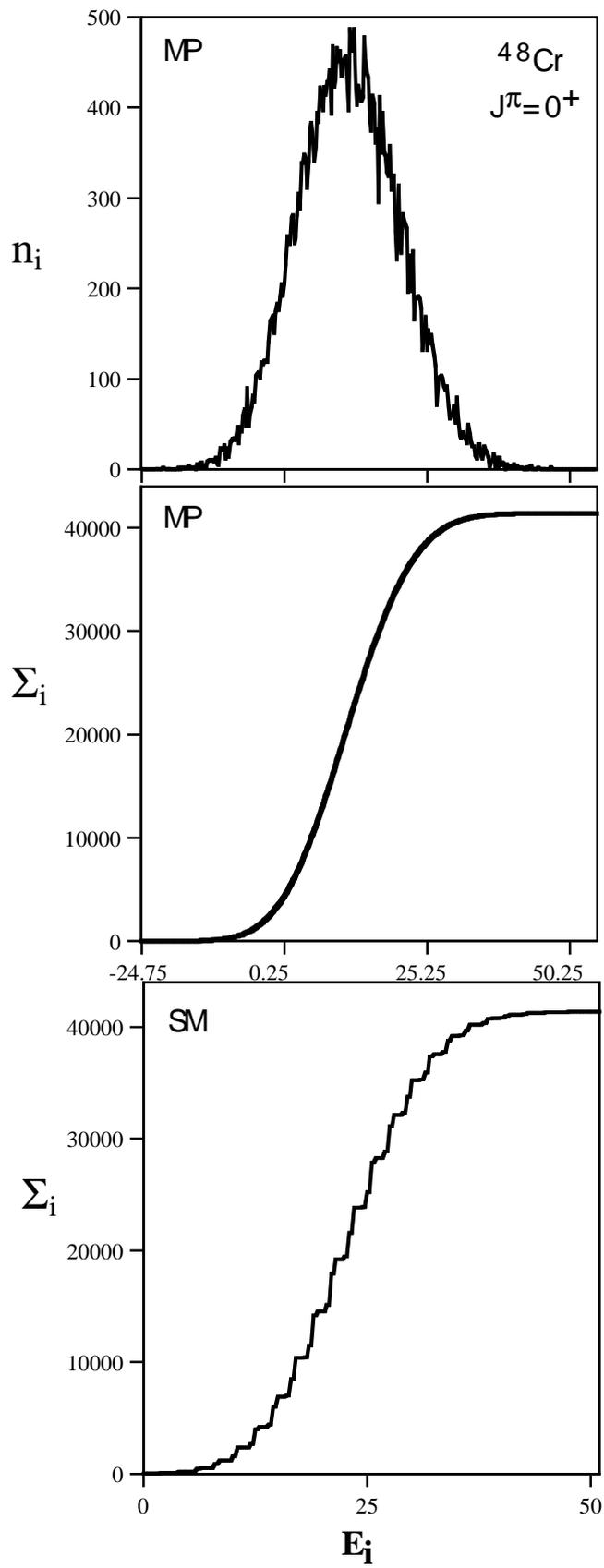
\begin{figure}
\caption{\label{fig2} Energy distribution of the $J^\pi = 0^+$
basis states in $^{48}$Cr obtained by the multipartition  method
(MP) and corresponding running sum. This  is  compared with the
running sum of the standard shell model (SM) basis states.}
\end{figure}

For illustrative purposes we discuss only few of the lowest states
of the nuclei under investigation. As shown in the plot of Fig.
\ref{fig1}, the sampling parameter $\epsilon$ varies very smoothly
with the dimensions $n$ of the reduced matrices. In these, as in
all other states considered, it scales according to
\begin{equation}
\epsilon= b \frac{N}{n^{2}} \exp{\left[-c \frac{N}{n}\right]} 
\label{epsscale}
\end{equation}
In virtue of this law, the increment of the dimensions of the
matrix is modest for large values of $\epsilon$, but grows
dramatically as $\epsilon$ gets smaller and smaller. This behavior
reflects the density of the unperturbed levels which is very low
at low energy and raises steeply around a centroid at high energy.
This is shown in the upper panel of Fig. \ref{fig2}. It follows
that the running sum of the unperturbed basis states grows very
slowly at low energy, then raises steeply toward its saturating
full $N$ value in a relatively small energy interval (middle panel
of Fig. \ref{fig2}). It is important to point out that this is the
distribution of  our unperturbed correlated states defined by Eq.
(\ref{part1}). The partitioning of the shell model space is
responsible for the squeezing of their energies around a centroid.
Had we adopted the standard SM basis, the state distribution would
have resulted more spread out and the running sum would have grown
discontinuously through several steps in correspondence of  each
sub-shell (lower panel of Fig. \ref{fig2}).
 
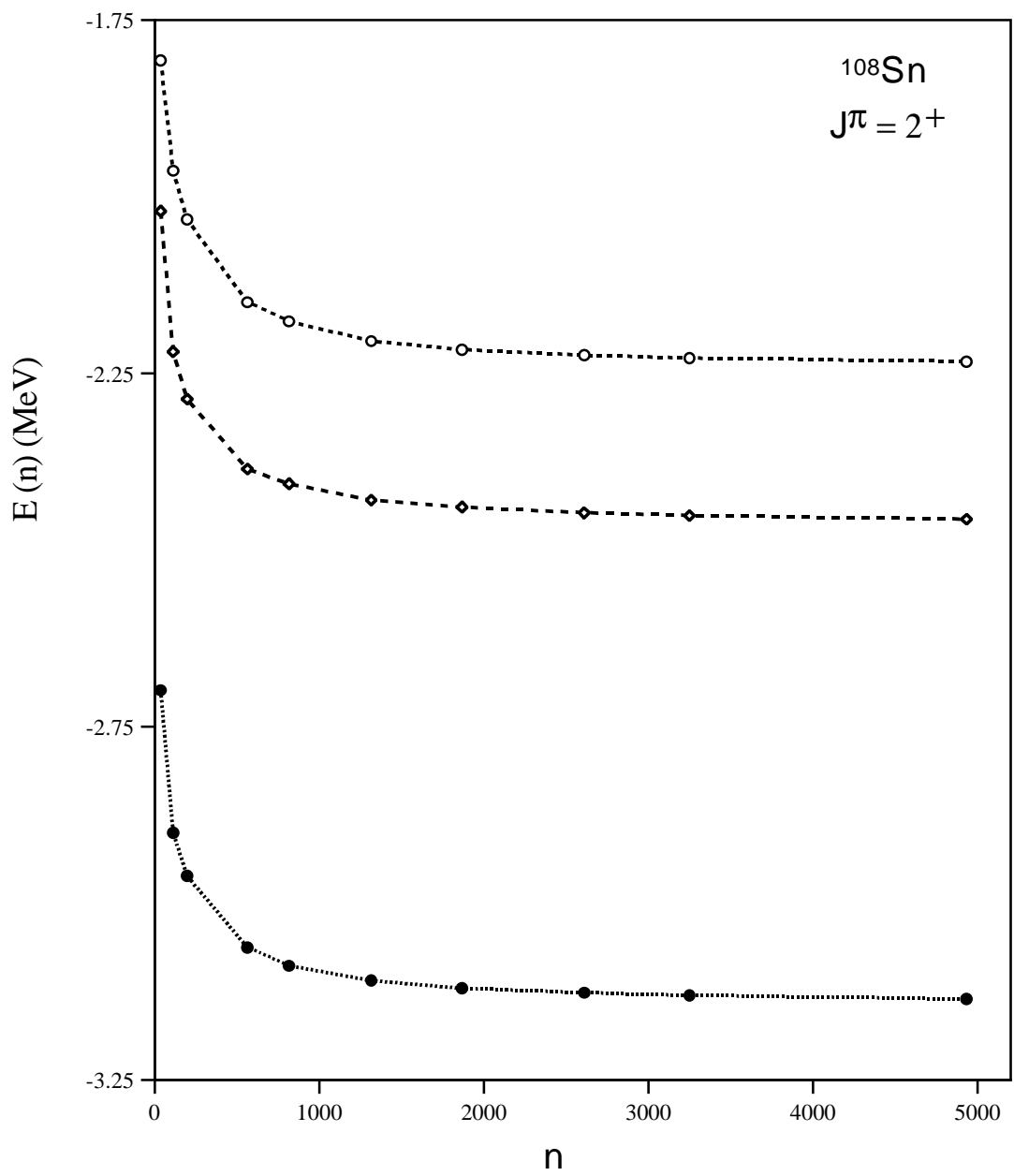
\begin{figure}
\caption{\label{fig3} low-lying $J^\pi = 2^+$ eigenvalues of
$^{108}$Sn versus the dimensions $n$ of the truncated matrices
resulting from the sampling.}
\end{figure}

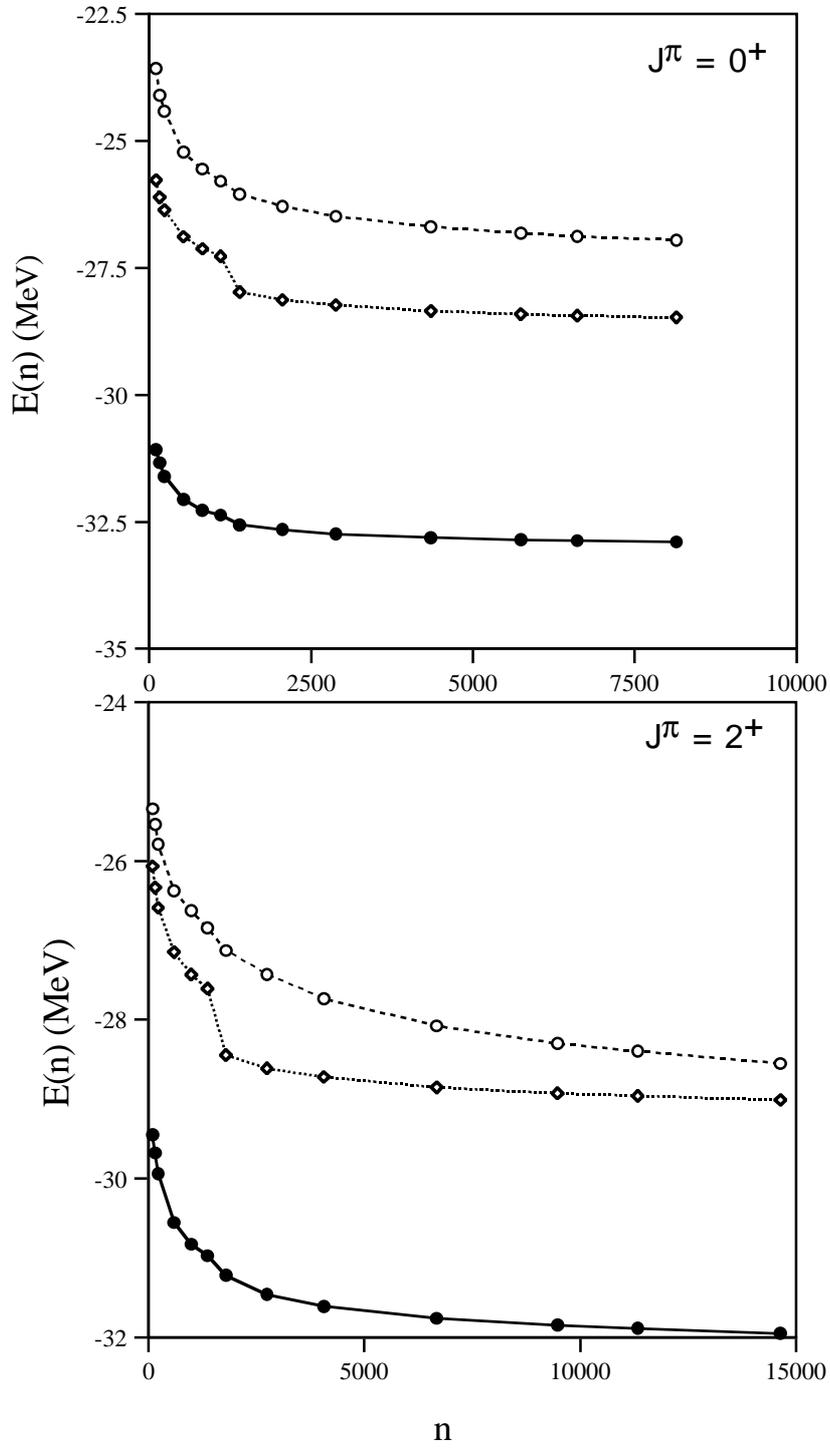
\begin{figure}
\caption{\label{fig4} The same as Fig. \ref{fig3} for $^{48}$Cr}.
\end{figure}

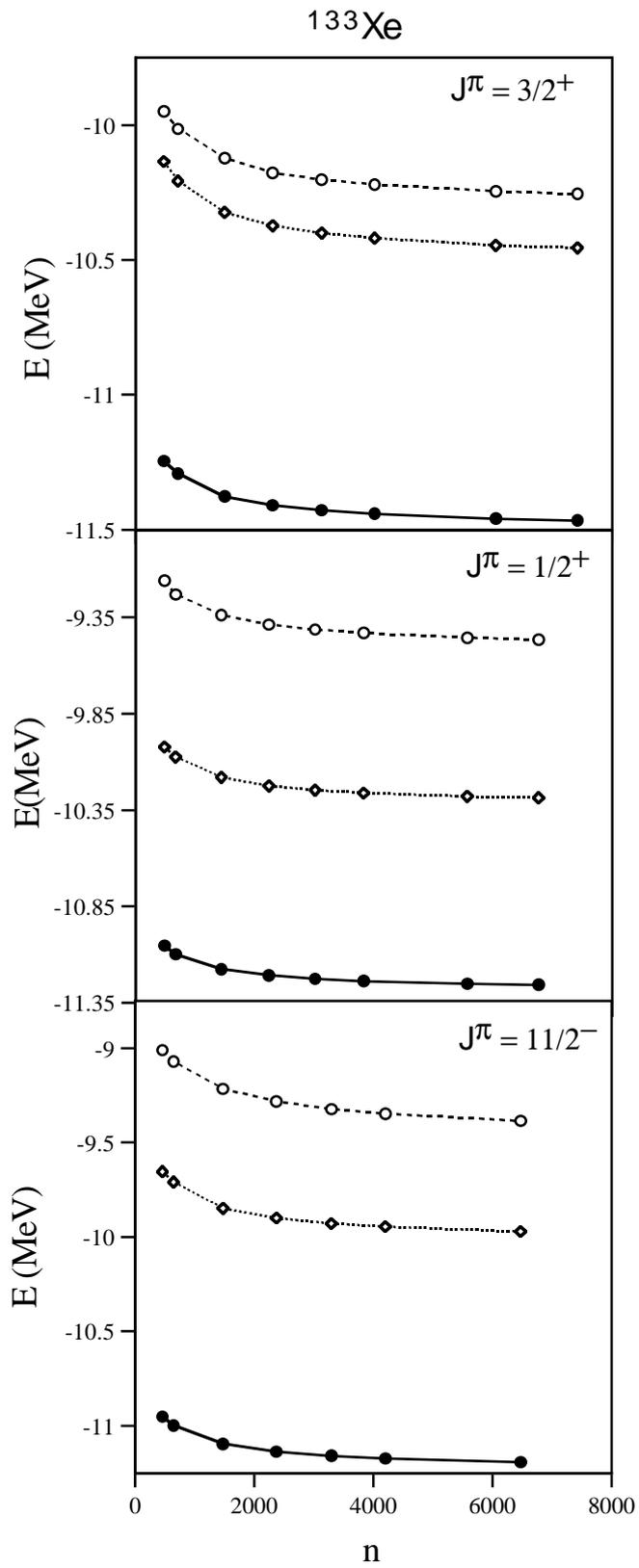
 \begin{figure}
\caption{\label{fig5} The same as Fig. \ref{fig3} for $^{133}$Xe.}
\end{figure}

Figs. \ref{fig3}-\ref{fig5} plot the eigenvalues versus the
dimensions $n$ of the matrices resulting from decreasing values of
the sampling parameter $\epsilon$ for some low-lying states of
$^{108}$Sn, $^{48}$Cr, and
 $^{133}$Xe, respectively. 
 
In all nuclei and for all states, the eigenvalues decrease
monotonically and smoothly with $n$. The only meaningful
exceptions are represented by the  curves of the first excited
$J^\pi = 0^+$ and $J^\pi = 2^+$ and few other,
similarly behaving, high-lying states of $^{48}$Cr. These undergo a jump from
an upper to a lower curve at  some small value of the sampling
parameter $\epsilon$, a signal of energy crossing.
Even in these cases, however, the subsequent behavior of the
energies is smooth as for the other
states. It follows that, in all cases, starting from a
sufficiently small $\epsilon$, the energies scale with the
dimensions $n$ according to the law
\begin{equation}
E=E_{0}+ b \frac{N}{n}\, \exp{\left[-c \frac{N}{n}\right]}
\label{scale}
\end{equation} 
where $b$, $c$, and $E_{0}$ are constants specific of each state
and the full dimension $N$ provides the scale. This fit allows for
an extrapolation to asymptotic eigenvalues which differ from the
exact ones in the second or third decimal digit. This is
explicitly proven for $^{108}$Sn and $^{48}$Cr (Table \ref{tab1}).
The reliability of the extrapolation in the case of $^{133}$Xe is
inferred from the rapid convergence of the iterative procedure, as
clearly illustrated in Fig. \ref{fig5}. Indeed,
the curves reach a plateau of practically constant 
energies starting from an $n$ value which  is smaller than the full 
dimension $N$ by
more than one order of magnitude in $^{108}$Sn, $^{48}$Cr, and
by more than two in $^{133}$Xe. 

Our exponential
extrapolation law is somewhat different from the one proposed in
Refs. \cite{Horo99,Horo03}. On the other hand, it is valid for all
states and nuclei examined and follows directly from the sampling,
as it can be inferred from the following heuristic argument.
 
Let us consider  the simplest case of one-dimensional eigenspace
($v=1$). From Eqs. (\ref{disprel}) and (\ref{samp1}), we can write
the corresponding sampling prescription as
\begin{equation}
\Delta \lambda =  \sum_j \Delta \lambda_j = \sum_j
(\lambda^{\prime}_j - \lambda_j) = - \sum_j
\frac{b^{2}_{1j}}{a_{jj} - \lambda_j - \Delta \lambda_j }.
\label{disprel1}
\end{equation}
Expanding $\Delta \lambda_j$, we get a solution whose leading term
is
\begin{equation}
\Delta \lambda_j =  \frac{b^{2}_{1j}}{a_{jj} - \lambda}.
\label{pert}
\end{equation}
>From the plots it is clear that the extrapolation law holds for an
energy range of 1-2 MeV in proximity of the exact eigenvalue. It
accounts therefore for small contributions coming from a small
fraction of the basis states in the $\sum_j$ in the high density
region around the peak, as shown in Fig. \ref{fig2}. Since, in
this range, the energies $E_j$ of our correlated basis states are
closely packed and approach the $E_n$ value, we can put $(a_{jj} -
\lambda)\propto n$ in first approximation. It remains to analyze
the matrix elements $b^{2}_{1j}$. These are given by
\begin{equation}
\langle \phi_j \mid v \mid j \rangle^2  \label{b1j}
\end{equation}
where
\begin{equation}
\mid \phi_j \rangle  = \sum_{i=1}^{j} c_{i}^{j} \mid i \rangle
\label{fij}.
\end{equation}
For the lowest eigenvalue, the dominant $c_{i}$ components of
$\mid \phi_j \rangle$ are the low-energy ones with small $i$
values. It follows that the  probability, $b^{2}_{1j}$, that the
interaction couples $\mid \phi_j \rangle$ to $\mid j \rangle$ is
small and random for $j \leq n$ and vanishes for $j>n$, as
prescribed by the sampling criterion. This implies that the
dimension $n$  represents the range of the allowed {\it events}.
We can therefore put $b^{2}_{1j} \propto exp (-\bar{j} /n)$, where
$\bar{j}$ is a label representative of the small fraction of $j$
terms  of the sum $\sum_j$. We used the factor $N$ to fix the
scale. The scaling law (\ref{epsscale}) for $\epsilon$ follows from the one 
for the energy $E$ since $\epsilon$ is essentially the derivative 
of $E$ with respect to $n$.   
 
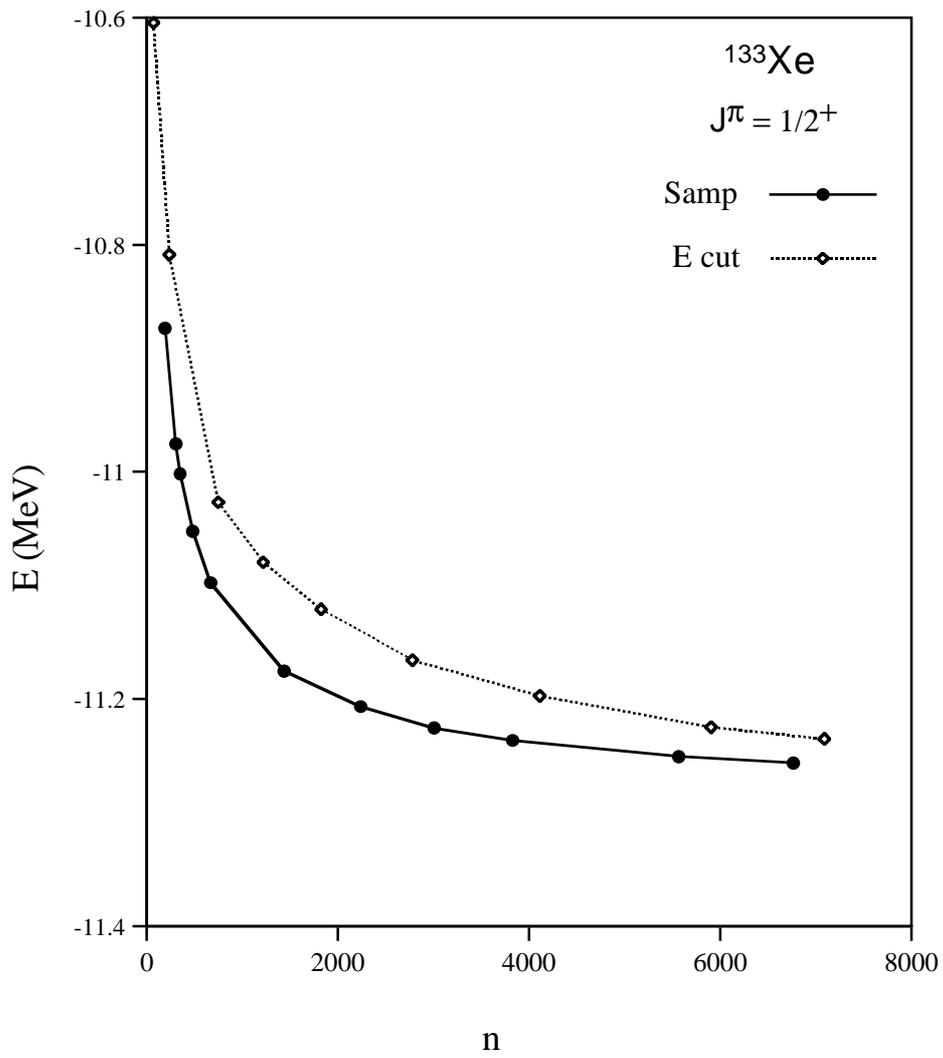
\begin{figure}
\caption{\label{fig6}  Eigenvalues obtained by the sampling and by
energy truncation versus the truncated dimensions $n$.}
\end{figure}

We have also compared the sampling with the energy truncation of
the Hamiltonian matrix. As shown in Fig. \ref{fig6}, the sampling
is obviously more effective and accurate. The two procedures,
however, tend to become equivalent for a value of $n$ which,
though large, is still much smaller than the full dimension $N$.
It is to be pointed out, however, that, in our case, the
effectiveness of the energy truncation  is due to a great extent
to the use of the correlated basis obtained through the
multi-partitioning method.

\subsection{Eigenvectors and E2 transitions}

An accuracy of the same quality is reached for the eigenstates
of the $n$-dimensional truncated Hamiltonian matrix
\begin{equation}
\psi_n = \sum_{i=1}^{n} c_{i}^{(n)} \mid i \rangle,
\end{equation}
where $\mid i \rangle$ are the correlated basis states obtained by
the partitioning method.

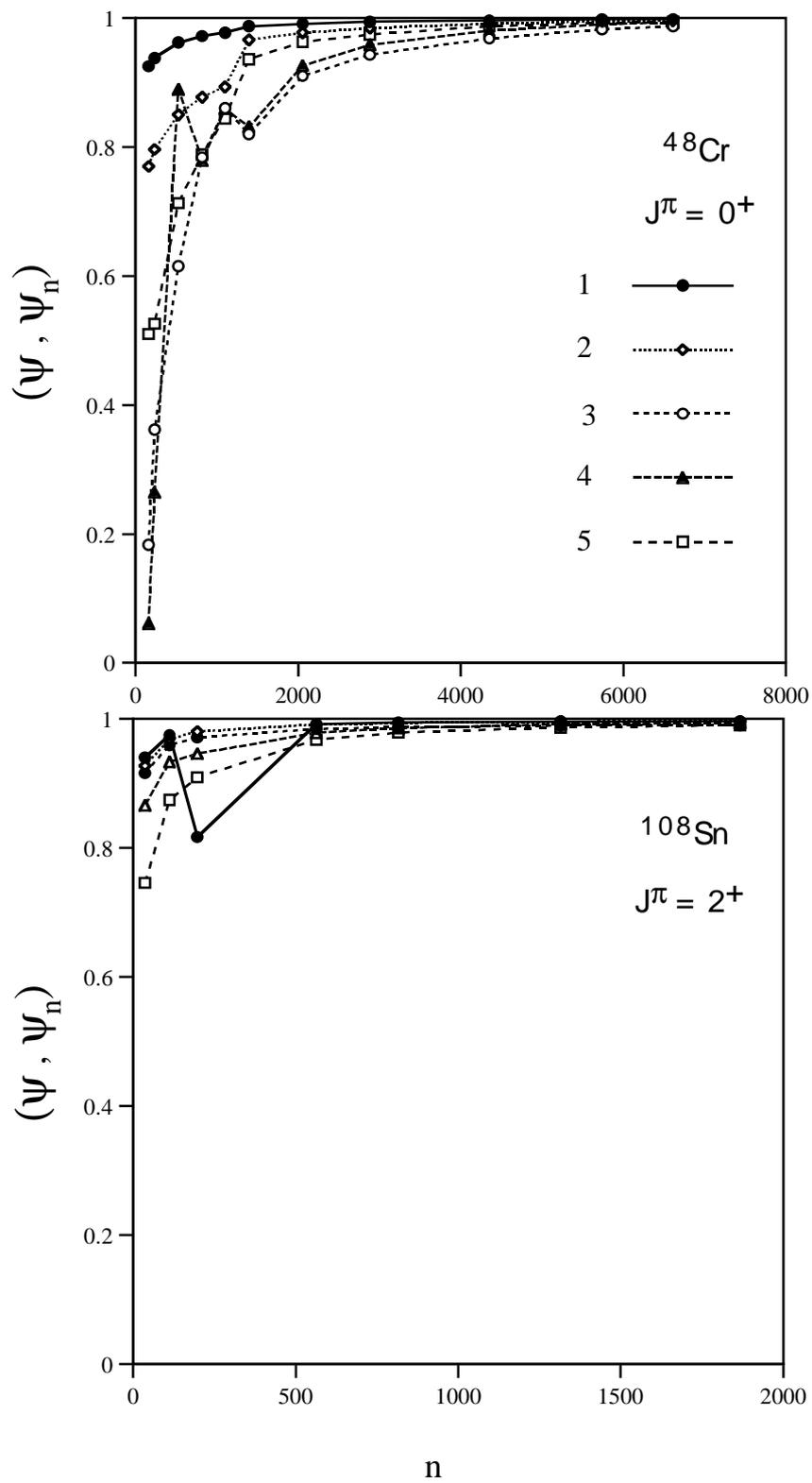
\begin{figure}
\caption{\label{fig7} Overlap of the lowest five $J^{\pi} =2^{+}$
and $J^{\pi} =0^{+}$ sampled eigenfunctions  with  the
corresponding exact ones in $^{108}$Sn and $^{48}$Cr,
respectively.}
\end{figure}

In Fig. \ref{fig7} we give the the overlap of $\psi_n$ with the
exact eigenvector $\psi$  for the first five $J^{\pi}= 2^{+}$
states of $^{108}$Sn and $J^{\pi}= 0^{+}$ of $^{48}$Cr. The
convergence to unity is fast for all five states, even if, for
some of them,  the overlap is very small at small $n$. Small
fluctuations are noticeable at small n values. They reflect the
interference  between the components of different wave functions
in correspondence of partial energy crossings. The above two
features represent a further proof of the {\it robustness} of the
iterative algorithm.

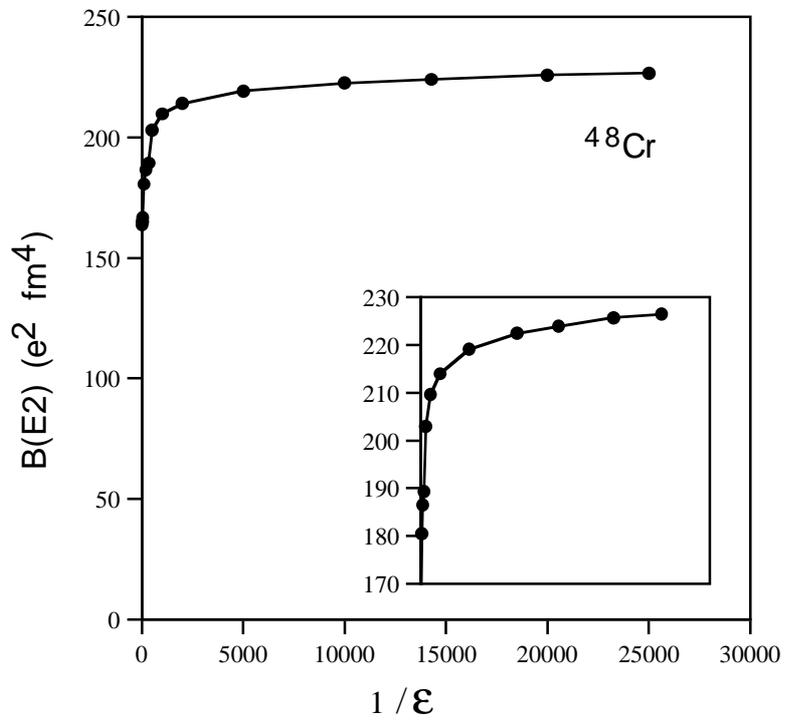
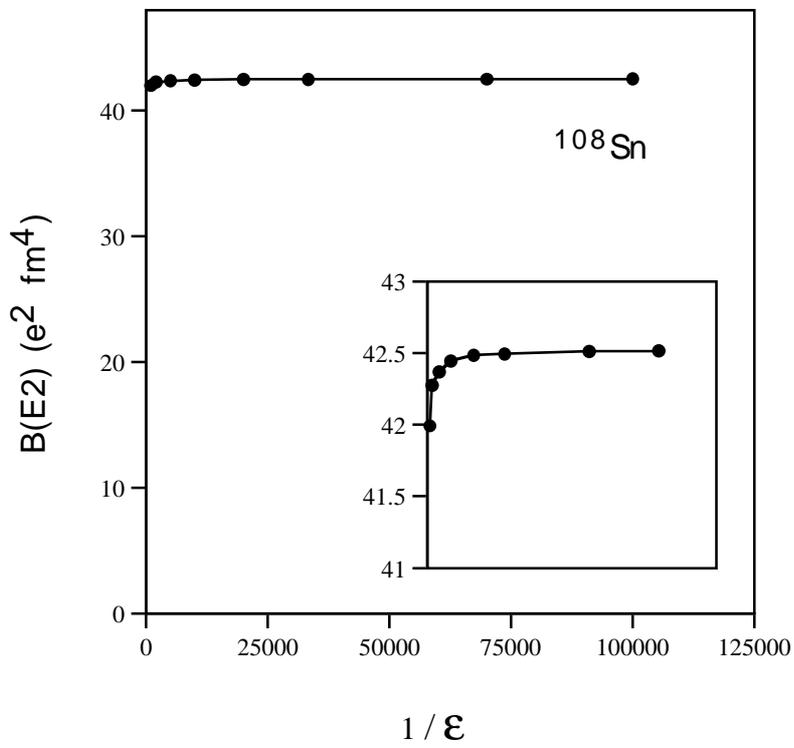
\begin{figure}
\caption{\label{fig8} Sampling of the strengths of  $J^{\pi}
=2^{+} \rightarrow J^{\pi} =0^{+}$ $E2$ transitions in $^{108}$Sn
and $^{48}$Cr.}
\end{figure}

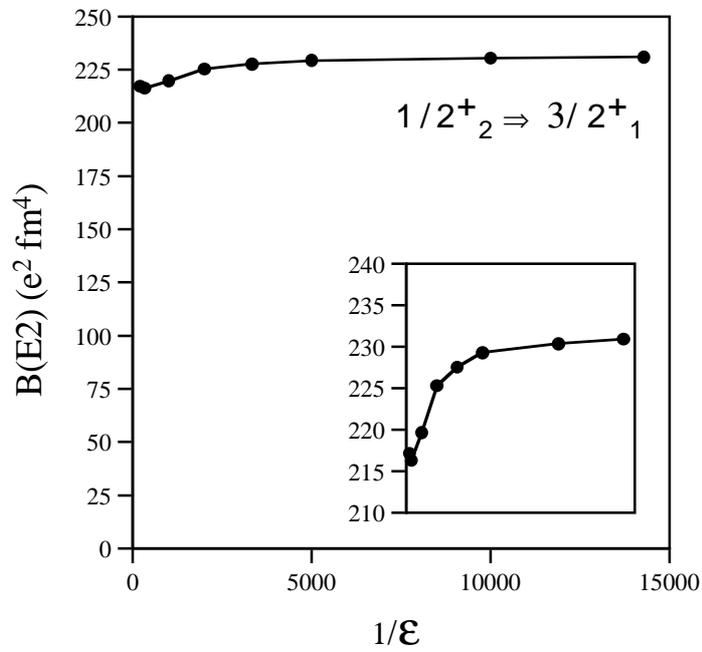
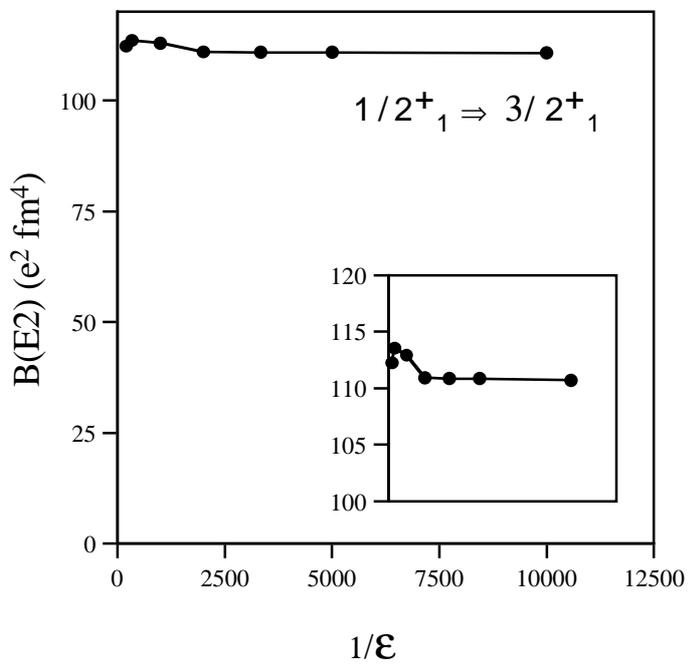
\begin{figure}
\caption{\label{fig9} Sampling of the strengths of  some $E2$
transitions in $^{133}$Xe.}
\end{figure}

To complete the analysis we  study the behavior versus $1/\epsilon
$ of the strengths of the $E2$ transitions between some low-lying
states in $^{108}$Sn and $^{48}$Cr ( Fig. \ref{fig8} ), as well as
in $^{133}$Xe (Fig. \ref{fig9}).  In all cases, the strengths
reach soon a plateau and, then, undergo very small variations,
appreciable only on a very small scale (see inset).  This
fine tuning analysis shows that each strength grows slowly with
$1/\epsilon$, apart from the transition $1/2^{+}_{1} \rightarrow
3/2^{+}_{1} $, whose strength decreases at an even slower rate. In  
all cases, their smooth behavior allows for an 
extrapolation to asymptotic values through a formula having the same 
structure as the  scaling law adopted for the energies (Eq. \ref{scale}).
Table \ref{tab2} shows that the strengths
computed at a relatively large $\epsilon$ differ very little from
the extrapolated ones, which in turn practically coincide with the
exact values. This rapid convergence is quite significant in view
of the extreme sensitivity of the transition strengths to even
very small components of the wave function.

\section{Concluding remarks}

We have shown that the importance sampling is inherent in the
iterative algorithm for diagonalizing large matrices. The
truncation of the dimensions of the matrices it promotes  is not
only kept under strict control, but is also quite severe when the
eigenvectors are highly localized.  For shell model nuclear
Hamiltonians, we achieved this localization by adopting a
correlated basis obtained by partitioning the shell model space
into two or more subspaces. As clearly illustrated by the
calculations carried out on some typical nuclei, the  sampling so
implemented allows to reduce the sizes of the Hamiltonian matrix
by at least an order of magnitude with no detriment of the
accuracy. We get in fact very accurate energies, wave functions
and $E2$ reduced strengths. Moreover, it generates extrapolation
laws to asymptotic eigenvalues and $E2$ transition probabilities which 
coincide practically with the exact corresponding quantities, whenever 
these are available.

It is important to point out that the method is specially
effective  when applied to $^{133}$Xe, having a neutron excess. 
On the ground of this result, we
feel confident that the sampling, combined with the use of the
correlated basis, will enable us to face successfully the
eigenvalue problem in heavier nuclei, all having a neutron excess.  
We also like to stress that the partition method is
specially suitable for enlarging the standard shell model valence
space through the inclusion of n-particle n-hole correlated
configurations. We plan to make such an implementation in order to
study the intruder states in light as well as heavy nuclei.

\begin{acknowledgements} The work was partly supported by the Prin
01 of the Italian MURST
\end{acknowledgements}

\newpage

\newpage

\begin{table}
\caption{\label{tab1} Approximate, extrapolated and exact (when
available) energies (in MeV).The extrapolation parameters are also
given.}
\begin{tabular}{cccccccccc}
 $^AX$ & $J^\pi_i$& N& $E_0$&b&c & n&$ E_i^{(n)}$ & $ E_i^{(extr)}$ & $E_i{(ex)}$\\
\hline $^{108}$Sn& $2^+_1$ & 17467&-3.145&0.0028&0.0032&1867
&-3.120 &-3.143(001) & -3.151 \\
 & $2^+_2$ & &-2.466&0.0029&0.0034 &&-2.439 & -2.463(001) &  -2.469 \\
 & $2^+_3$ & &-2.244&0.0030&0.0021 & &-2.216& -2.244(001) & -2.266  \\
\hline$^{48}$Cr &  $0^+_1$& 41355 &-32.972&0.01721&0.00456
&5739&-32.851&-32.955(019) &-32.954  \\
  &  $0^+_2$&  &-28.638 & 0.03450&0.01486&&-28.414&-28.604(011)&-28.565   \\
  &  $0^+_3$& & -27.166&0.06038&0.01280 & &-26.819&-27.166(042)& -27.158  \\
    &  $2^+_1$&182421  & -32.164&0.019&0.008 &14642&-31.952&-32.145(020) & 
    -32.148\tablenotemark[1]\\
    &  $2^+_2$& &-29.133&0.0114&0.0009 &&-29.012 & -29.122(016)&   \\
    &  $2^+_3$&  &-28.745 &0.0238&0.0037& &-28.552&-28.721(053)&   \\
\hline$^{133}$Xe &  $1/2^+_1$& 125756 &-11.280&0.0014&0.0018 &6763&-11.256&-11.279(002) &  \\
&  $1/2^+_1$&  &-10.319&0.0019& 0.0019& &-10.285 &-10.317(001) &  \\
&  $1/2^+_2$& &-9.498&0.0021& 0.0019 & &-9.464&-9.496(007) &  \\
&  $3/2^+_1$&242308 &-11.491&0.0009&0.0012 &7417&-11.466&-11.490(002)&   \\
&  $3/2^+_2$&  &-10.491&0.0012& 0.0012 &  &-10.455&-10.490(003) &\\
&  $3/2^+_3$& &-10.292&0.0013& 0.0013 &&-10.255&-10.292(002) &  \\
&  $11/2^-_1$&497666 &-11.222&0.0004& 0.0005&6472&-11.192&-11.222(003) &  \\
&  $11/2^-_2$&   &-10.013&0.0006&0.0005 &&-9.971&-10.012 (003)  &  \\
&  $11/2^-_3$&    &-9.439&0.0008&0.0002 &&-9.386 &-9.438 (008)  & \\
\end{tabular}
\tablenotetext[1]{from Ref.\ \cite{Brand98}.}
\end{table}

\begin{table}
\caption{\label{tab2} Approximate, extrapolated and exact (when
available) $B(E2)$ (in $e^2 fm^4$).}
\begin{tabular}{cccccccc}
 $^AX$ & $J^\pi_i \rightarrow J^\pi_f$&$\epsilon$&$n_i$&$n_f$&$ B(E2)_{(\epsilon)}$ &$ B(E2)_{(extr)}$&$ B(E2)_{(ex)} $ \\
\hline $^{108}$Sn& $2^+_1 \rightarrow 0^+_1$ &
1.x10$^{-4}$&1867&1034&42.45& 42.68(02) &  42.71  \\
\hline$^{48}$Cr &   $2^+_1 \rightarrow 0^+_1$& 4.x10$^{-5}$&14642&8144&226.5&227.8(8) &  228  \\
\hline$^{133}$Xe &  $1/2^+_1 \rightarrow 3/2^+_1$&7.x
10$^{-5}$&6763&7417&110.1&109.3(7)  &    \\
 &  $1/2^+_2 \rightarrow 3/2^+_1$&  &&&230.9&233.0(3) &    \\
\end{tabular}
\end{table}

\end{document}